\documentclass{article}
\begin{document}
\baselineskip18pt

\title{A Subtlety of the Schr\"{o}dinger Picture Dynamics}
\author{Piotr Garbaczewski\thanks{p.garbaczewski@if.uz.zgora.pl}\\
 Institute of Physics,  University
of Zielona G\'{o}ra,  Poland}
\maketitle
\begin{abstract}
We address  a mathematical and physical status of exotic (like
e.g. fractal) wave packets and their quantum dynamics. To this
end, we  extend the formal  meaning of the Schr\"{o}dinger
equation  beyond the domain of the Hamiltonian. The dynamical
importance of the finite mean energy condition is elucidated.
\end{abstract}

 \vskip0.3cm
\section{Motivation}
For a simple Hamiltonian system whose energy operator $\hat{H}$
has a countably infinite  spectrum $\{ E_n\}$ and normalized
eigenvectors $\{ |n\rangle \} $, so that   $\hat{H}|n\rangle =
E_n|n\rangle
 $ and  $\langle m|n\rangle = \delta _{mn}$, any   general  pure  state  of the system
 is  defined in terms  of  a normalized superposition $|\psi ,0\rangle \doteq \sum c_n |n\rangle $,
  $\sum |c_n|^2 =1$.
 Its  unitary  time evolution  follows:
\begin{equation}
 |\psi ,t\rangle = \sum c_n \exp \left(- {\frac{iE_nt}{\hbar }}\right)  \,
 |n\rangle \doteq \exp \left(- {\frac{i\hat{H}t}{\hbar }}\right) |\psi
 ,0\rangle \, .
 \end{equation}

Since $\hat{H}$  typically is an unbounded Hilbert space operator,
we need  a number of precautions  concerning its domain
properties,  to infer  the Schr\"{o}dinger  equation
\begin{equation}
i\hbar {\frac{\partial }{\partial t}} |\psi ,t\rangle = \hat{H}  \label{one}
|\psi , t\rangle
\end{equation}
while it is often \cite{hall}  taken for granted,  that  the operator  identity
\begin{equation}
\left(\hat{H}  - i\hbar {\frac{\partial }{\partial t}}\right)
|\psi ,t\rangle  =0   \label{two}
\end{equation}
unquestionably holds true for  all states of the system, even most
exotic like e.g. fractal wave packets of
Refs.~\cite{birula}-\cite{hall}.

The two  manifestations, (\ref{one}) and  (\ref{two}), of the
Schr\"{o}dinger picture quantum dynamics are  inequivalent for all
states which are not in the domain of the Hamiltonian. A
particular example of this situation is provided  in
\cite{birula}: the  time evolution of a fractal  quantum state,
represented by a continuous but nowhere differentiable function
$|\psi ,t\rangle $,  can per force   be  related to the
Schr\"{o}dinger equation, but   in  a   weak sense, by demanding
that:
\begin{equation}
E_n \langle n|\psi ,t\rangle  = i\hbar  {\frac{\partial }{\partial
t}} \langle n|\psi ,t\rangle
 \label{three}
\end{equation}
hold true for all $n$.  The left-hand-side may be interpreted as a
scalar product of two  Hilbert space vectors
 $\hat{H} |n\rangle$ and $|\psi ,t\rangle$, but surely cannot be rewritten as
 $\langle n|\hat{H}|\psi ,t\rangle $.

The present paper is devoted to a deeper  discussion of  deceivingly simple formulas
 (\ref{one}), (\ref{two}), (\ref{three}) and their mutual relationships.  Since domain problems
 are encountered  while evaluating mean values of unbounded observables, we pay particular
attention to the importance  of the finiteness of  mean  energy
condition.

If this restriction is violated,  one encounters "infinite mean
energy" states, \cite{birula,hall}.  On the physical grounds they
are irrelevant (nonexistent), since one would need an infinite
energy to create (prepare) them.  As well, by considering such
states  as a limiting case of an approximation procedure, in terms
of a sequence  of states with increasing finite energy, one ends
up with a standard mathematical non-existence problem   for the
mean value.

The finite mean energy  condition is  known to be  an important
technical input that entails  a  trajectory
 interpretation of the Schr\"{o}dinger picture quantum  dynamics,
  in terms of sample paths of a  Markov diffusion-type process, \cite{carlen} .
Since there appeared published claims, \cite{hall}, that standard
trajectory interpretations fail for a certain class of wave
functions that have well defined quantum evolutions, we indicate
why evolutions considered in \cite{hall} are in   fact
ill-defined.

\section{Schr\"{o}dinger equation}

 Let  ${\cal{U}}(\cal{H})$ denote a family of functions of a real variable $t \in R$ with values in
  ${\cal{H}}= L^2(R^n;dx)$, such that:\\
  (i) functions $\psi (t)$   are continuous i.e. $\lim_{t\rightarrow t_0} \| \psi (t) - \psi (t_0)\| = 0$\\
  (ii) we  have  $\| \psi (t_1)\| = \| \psi (t_2)\| $ for all $t_1, t_2 \in R$.\\

A function $\psi (t)$ is called strongly differentiable if for
each  value  $t\in R$  there exists $\psi '(t) \in \cal{H}$
obeying:
\begin{equation}
\lim _{t' \rightarrow t} \| {\frac{\psi (t') - \psi (t)}{t'-t}} - \psi '(t)\| = 0 \, .
\end{equation}
Then we write $\psi '(t) = {\frac{d}{dt}} \psi (t)$.

Let $\hat{H}$ be a self-adjoint operator with the  dense  domain
$D(\hat{H}) \subset \cal{H}$.  We say that $\psi (t)$ obeys the
Schr\"{o}dinger equation (set $\hbar =1$)
\begin{equation}
\hat{H} \psi  (t) = i {\frac{d\psi (t)}{dt}}  \label{schr}
\end{equation}
if the following  three conditions  are valid: \\
(a) $\psi (t) \in D(\hat{H})$ for all $t \in R$,\\
(b) $\psi (t)$ is strongly differentiable,\\
(c) the equality  in  Eq.~(\ref{schr}) is verified to  hold true.\\

Let us notice that  to handle the left-hand-side of (\ref{schr}) the condition (a)  is both necessary and
sufficient. As far as the right-hand-side is concerned, we only  need to know that $\psi (t)$ is strongly differentiable,
which has nothing in common with the condition (a).
Once we have a strongly differentiable function $\psi (t)$ for which - additionally - (a) holds true,   we
are ultimately allowed to check (c). If so, we can tell that $\psi (t)$ actually is a solution
of the  Schr\"{o}dinger  equation.

Let us denote $E_{\lambda } , \lambda \in R$  a resolution of
unity for    $\hat{H}$ i.e.    $\hat{H} = \int_{-\infty }^{+\infty } \lambda
dE_{\lambda }$. If $\phi $ is a continuous function of a real
variable (continuity is presumed for convenience, but it is not  a
must), then we define a "function of an operator"  $\phi (\hat{H})
= \int_{-\infty }^{+\infty } \phi (\lambda ) dE_{\lambda }$. Its
domain is $\{ f\in {\cal{H}};
 \int_{-\infty }^{+\infty } |\phi (\lambda )|^2 d(f,E_{\lambda }f) < \infty  \}$.

If $\phi $ is bounded, then $\phi (\hat{H})$  is a bounded operator.
 In particular,  if we take $a<b $ in $R$ and
 $\psi \in {\cal{H}}$, then the function
 \begin{equation}
 \psi _{a,b}(t) = \int_a^b \exp(-i\lambda t) d(E_{\lambda }\psi ) = \int_{-\infty }^{+\infty } \exp (-i\lambda t)
 d(E_{\lambda }[E_b - E_a]\psi )  =  \exp(-\hat{H}t) [E_b - E_a]\psi
 \end{equation}
 fulfills conditions (a),  (b) and (c),  that is  solves the  Schr\"{o}dinger  equation.

We have  a strong convergence  $\lim_{b\rightarrow +\infty , a\rightarrow -\infty } [E_b - E_a] \psi $ and also
\begin{equation}
\lim_{b\rightarrow +\infty , a\rightarrow -\infty } \psi _{a,b}(t) = \int_{-\infty }^{+\infty }  \exp(-i\lambda t)
d(E_{\lambda } \psi )  = \exp(-i\hat{H}t) \psi = \psi (t)
\end{equation}
where however   $\psi (t)$ needs not to belong to the domain od $\hat{H}$. In such case  one cannot
even attempt to  verify  the equality (\ref{schr}).  Nonetheless $\exp(-i\hat{H}t) \psi = \psi (t)$ is
well defined.

\section{Quantum evolution beyond the domain of $\hat{H}$}

We shall give a rigorous meaning to the formula (\ref{two}) when extended to functions not
 belonging to the domain of $\hat{H}$.   Let ${\cal{E}}$ denote  a family of continuous
 functions  $u$  of a real variable,  taking values in $R$ and
 such that  $\lambda -   u(\lambda )$ is a bounded function.   By ${\cal{E}}_{\hat{H}}$  we denote a
 family of  functions  with values in the Hilbert  space ${\cal{H}}$ defined as follows:
 \begin{equation}
 {\cal{E}}_{\hat{H}} = \{\,  \psi ^u(t) = \int_{-\infty }^{+\infty } \exp [iu(\lambda )t]
 d(E_{\lambda } \psi )=  \exp[-iu(\hat{H})t] \psi \,  \}
 \end{equation}
 where $u\in {\cal{E}}$, $\psi \in {\cal{H}}$.  Clearly, ${\cal{E}}_{\hat{H}} \subset {\cal{U}}({\cal{H}})$.

Some care is needed to extend the formula   $(\ref{two})$ beyond the domain of $\hat{H}$.
To this end,  let us define an operation $\hat{S}$  with the domain ${\cal{E}}_{\hat{H}}$ and values in
${\cal{U}}({\cal{H}})$, which will be a  extension  of $(\hat{H} - i{\frac{d}{dt}})$.
 We take $\psi \in {\cal{H}}$ and introduce
$\psi ^u_{a,b}(t) = \int_{-\infty }^{+\infty } \exp[-iu(\lambda
)t] d(E_{\lambda }[E_b - E_a]\psi )$. Then $\psi ^u_{a,b}(t) \in
{\cal{E}}_{\hat{H}}$ and  obeys  both (a)  and (b).  Consequently,
$(\hat{H} - i{\frac{d}{dt}}) \psi ^u_{a,b}(t)$ is well defined and
we have
\begin{equation}
(\hat{H} - i{\frac{d}{dt}}) \psi ^u_{a,b}(t) = \int_{-\infty
}^{+\infty } [\lambda - u(\lambda )] \exp[-iu(\lambda )t]
d(E_{\lambda }[E_b - E_a]\psi )\, .
\end{equation}

Because of
\begin{equation}
 \|(\hat{H} - i{\frac{d}{dt}}) \psi ^u_{a,b}(t)\| ^2 \leq
 \int_{-\infty }^{+\infty } [\lambda - u(\lambda )]^2
 d(\psi ,E_{\lambda }\psi )
 \leq M \int_{-\infty }^{+\infty } d(\psi , E_{\lambda }\psi ) = M\| \psi \| ^2
 \end{equation}
 where $M= \sup_{\lambda } |\lambda - u(\lambda )|^2$,   it is possible  to extend
 $(\hat{H} - i{\frac{d}{dt}})$
 to an operator  $\hat{S}$ whose action on a function
 $\psi ^u(t) = \int_{-\infty }^{+\infty } \exp[- iu(\lambda )t] d(E_{\lambda }\psi )$
  reads as follows:
 \begin{equation}
\hat{S} \psi ^u(t) =\int_{-\infty }^{+\infty }  [\lambda - u(\lambda
)] \exp [- iu(\lambda )t]
 d(E_{\lambda }\psi ) \, .~\label{ext}
\end{equation}
Indeed,   we infer that $(\hat{H} - i{\frac{d}{dt}}) \psi ^u_{a,b}(t)$  converges strongly in ${\cal{H}}$
and uniformly in $t$ to $\int_{-\infty }^{+\infty }  [\lambda - u(\lambda
)] \exp [- iu(\lambda )t]
 d(E_{\lambda }\psi )$.  Thus,  we can define
 $\hat{S} \psi ^u_{a,b}(t)=(\hat{H} - i{\frac{d}{dt}}) \psi ^u_{a,b}(t)$ and  in view of Eq.~(\ref{ext})  the
  operator  $\hat{S}$ is an extension of $(\hat{H} - i{\frac{d}{dt}})$ to the whole of ${\cal{E}}_{\hat{H}}$.

In connection with the above extension  notion, let us  recall
that if we have operators $A$ and $B$ defined on their respective
domains $Q_A$ and $Q_B$ such that $Q_A \subset Q_B$ and if an
operator $B$ is defined on $Q_B$ so that $Ax=Bx$ for all $x\in
Q_A$,  then we  call $B$ an extension of $A$ and denote $A\subset
B$.  We actually  have $(\hat{H} - i{\frac{d}{dt}}) \subset
\hat{S}$.

At this point, we may consider
\begin{equation}
\hat{S} \psi (t) =0  \label{hall}
\end{equation}
which is clearly  an equation solved   by any function  $\psi ^u(t)$ with
$u(\lambda )= \lambda $.  This equation is a rigorous version of (\ref{two}), the fact which if often disregarded in the
literature, c.f. \cite{hall}. It is clear that (\ref{hall})  cannot be directly  rewritten in the form
(\ref{two}), unless with an  obvious  abuse of notation.

\section{Finite energy condition and  Hilbert space scale}

A necessary condition for the  equation (\ref{one}) to make sense
is the condition  denoted previously by (a):  $\psi (t) \in
D(\hat{H})$   for all $t$. We relate this property to  the
so-called finite mean energy condition,  which according to
\cite{carlen}
 is a limitation upon wave functions, necessary to ensure the existence
 of  a stochastic counterpart  of the  Schr\"{o}dinger picture evolution
 (i .e. well defined Markovian diffusion-type  processes).

We are here motivated by  \cite{birula,hall}.
The statement of Ref. \cite{hall} is: there exist pure quantum
states for which the mean energy is finite, but no consistent
Schr\"{o}dinger  evolution (\ref{one}) can be defined (in fact,
 our condition (a) does not hold true). A complementary statement of \cite{hall}
 and \cite{birula} is that: there exist wave functions which "have infinite mean energy".

In contrast to the reasoning of Ref. \cite{hall},  in \cite{birula} the pertinent
(fractal)  wave function is derived in a controlled way, through
   a well defined limiting procedure.  Since the  mean energy diverges in this limit,
   it is more correct to say  about the "nonexistence" of the mean value,  instead of
  invoking  a state with an "infinite mean energy".
  Let us discuss in some detail  the background and validity of these  claims.

We   assume that $\hat{H}$ is defined in ${\cal{H}}$ and is:
 (i) self-adjoint, (ii) is bounded from below,
 (iii) is unbounded from above.  In view of (ii), we may  always   replace a  given  Hamiltonian by a strictly
 positive  operator, hence we assume: (iv) $\hat{H}$ is strictly  positive i. e.
 there is $m>0$ such that for all $\psi \in D(\hat{H})$  we have     $(\psi ,\hat{H}\psi )
 \geq m \|\psi \|^2 $.

The domain $D(\hat{H}) \subset {\cal{H}}$ is   a linear space with
the scalar product of ${\cal{H}}$. However, $D(\hat{H})$ is not  a
Hilbert space: $D(\hat{H}) \neq {\cal{H}}$ and $D(\hat{H})$ is
dense in ${\cal{H}}$.  Thence $D(\hat{H})$ is not complete,
\cite{glazman}.

In the linear space $D(\hat{H})$  we introduce a new scalar product:
\begin{equation}
(f,g)_2 \doteq  (\hat{H}f,\hat{H}g)
\end{equation}
where $f, g \in D(\hat{H})$ and $(\cdot ,\cdot )$ is the scalar
product in ${\cal{H}}$. Since  $\hat{H}$ is strictly positive, one
can prove that the   hitherto  incomplete  linear space
$D(\hat{H})$ becomes complete in the new norm inferred from
$(\cdot ,\cdot )_2$. With this scalar product $D(\hat{H})$
actually  \it is \rm  a Hilbert space which we denote   $ {\cal{H}}_2$. We
have the set inclusion  $ {\cal{H}}_2 \subset {\cal{H}}$ and $\|f\| \leq \|f||_2$
for all $f \in {\cal{H}}_2$.

We can define a number  of other scalar products on $D(\hat{H})$,
like e g.
\begin{equation}
(f,g)_k = (H^{k/2}f,H^{k/2}g)
\end{equation}
with $k=2,1,0,-1,-2$. The case of $k=2$ we have just  considered, while $k=0$ corresponds to the standard
Hilbert space scalar product in ${\cal{H}}$.  Each of these scalar products  defines a corresponding norm in
$D(\hat{H})$ according to $\| f\| _k = \| {\hat{H}}^{k/2}f\|$, for  $f\in D(\hat{H})$.

Let us stress that  $D(\hat{H})$   is complete
exclusively in the norm $\| \cdot \|_2$.   However, we  can
complete $D(\hat{H})$ to  respective  Hilbert spaces in each of the
considered norms, so arriving at the Hilbert space scale (the set  inclusion $\subset $ means also
$\neq $):
\begin{equation}
{\cal{H}}_2 \subset {\cal{H}}_1 \subset {\cal{H}}_0 ={\cal{H}} \subset {\cal{H}}_{-1} \subset {\cal{H}}_{-2}
\end{equation}
which is  parallelled  by a chain of  norm inequalities  $\|f\|_2
\geq \|f\|_1 \geq \|f\|_0=\|f\| \geq \|f\|_{-1} \geq \|f\|_{-2}$
for all $f\in D({H})\equiv{\cal{H}}_2$.

Let us consider ${\cal{H}}_1$ as a set of vectors which, by definition, contains $D(\hat{H})$ as a dense subset.
Therefore  for all $f \in D(\hat{H})$  we have:
\begin{equation}
\|f\|^2_1 = (f,f)_1 = (\hat{H}^{1/2}f,\hat{H}^{1/2}f) =
(f,\hat{H}f) \doteq  \langle \hat{H}\rangle _f \label{mean}
\end{equation}
where in addition   one can demonstrate that  ${\cal{H}}_1=
D({\hat{H}}^{1/2})$.

If $\psi  \in D(\hat{H})$, we traditionally  call $(\psi,
\hat{H}\psi )$ the mean energy  of the quantum system  in the pure
state $\psi $.  The mean value coincides with an ${\cal{H}}$-scalar product
of two legitimate Hilbert space vectors: $\psi $ and $\hat{H}\psi $.

Perhaps it is worthwhile to spell out the meaning of mean energy states $\psi $ which do not
belong to $D(\hat{H})$.
The previously mentioned claims of Ref. \cite{hall,birula}
appear to  ignore   the problem of how to handle the "mean
value" with $\psi $ which is  \it not \rm in the domain of
$\hat{H}$. The relevant statement in Sect. 3 of Ref.~\cite{hall}
 reads: "one can arrange $\hat{H}\psi $ to
diverge (as the series) almost everywhere, while keeping  the average energy
$\langle \hat{H} \rangle $ finite".

Let us come back to the formula (\ref{mean}). If $g \in
{\cal{H}}_1$ but $g$ is  \it not \rm  an element of $D(\hat{H})$,
it is not allowed to infer uncritically
 $\|g\|_1^2 = (g,\hat{H}g)$,  because $\hat{H}g$ is  not   defined.
 Nevertheless, since  we  have in hands a consistent definition of  $\|g\|_1^2 =
(\hat{H}^{1/2}g,\hat{H}^{1/2}g)$, in view of  Eq.~(\ref{mean})   the formula
$(\hat{H}^{1/2}g,\hat{H}^{1/2}g)$  may possibly  be interpreted as a straightforward  generalization  of
  the mean energy notion ($<\hat{H}>_g$, c.f.  Eq.~(\ref{mean})).
 Then   ${\cal{H}}_1$, with  ${\cal{H}}_2 \subset {\cal{H}}_1 \subset {\cal{H}}$, would stand
    for a natural extension   of the    set of  states with a finite mean energy
     beyond the domain  ${\cal{H}}_2$  of $\hat{H}$.

 For states beyond ${\cal{H}}_1$ we may     expect infinite values for $\|g\|_1^2$ to occur.
 Since $\|g\|_1^2$  is not defined, one may interpret any $g \in {\cal{H}}\backslash {\cal{H}}_1$ as
 an infinite energy state, in the sense that $<\hat{H}>_{E_a,g} = (E_ag,\hat{H}E_ag) \rightarrow \infty $
  as $a \rightarrow \infty $.

We have considered a selfadjoint, unbounded, strictly  positive
operator $\hat{H}$ in a Hilbert space $\cal{H}$.  $\hat{H}$ is
invertible and the inverse operator  $ \hat{H}^{-1}$ is  bounded
in $\cal{H}$.  Notice that for any $f\in {\cal{H}}$, we have
$\hat{H}^{-1}f \in D(\hat{H})$ and consequently:
\begin{equation}
\hat{H}\,  D(\hat{H})  = {\cal{H}}
\end{equation}
 and
\begin{equation}
\hat{H}^{-1}\,  {\cal{H}} =
D(\hat{H})\, .
\end{equation}

Therefore, for the operator $\hat{H}^{-1}$ we need not to bother about domain properties
and  for any $f \in \cal{H}$ we have the well defined mean value (a scalar
product of two Hilbert space vectors) $(f,\hat{H}^{-1}f)$.
Since any  $f\in {\cal{H}}$ can be represented in the form $f= \hat{H} \psi $ where
$\psi \in D(\hat{H})$, we have:
\begin{equation}
(f,\hat{H}^{-1}f)= (\psi , \hat{H}\psi ) \label{finite}
\end{equation}
which ultimately reduces  the finite mean  energy definition exclusively
to vectors from $D(\cal{H})$.  The mean energy notion appears not to have meaning beyond
$D(\hat{H})$, unless carefully generalized.

\section{Trajectory-based interpretations of quantum motion}

The original purpose of Ref.~\cite{hall} has been a critique  of "trajectory-based
interpretations of quantum mechanics" with two targets: Nelson's stochastic mechanics and
so-called Bohmian mechanics.
The point is that those two targets  refer  to the Schr\"{o}dinger picture  quantum
dynamics and not  the full-fledged formalism of quantum theory with varied experimental
connotations.

A well founded fact is that  at least
two different "trajectory pictures "  can be related to the very same mathematical model
 based on the Schr\"{o}dinger wave packet evolution:  deterministic  Bohmian
  paths  \cite{holland,durr}  and  random paths of
(basically singular)  diffusion-type processes,  \cite{carlen,eberle}.
Additionally, under suitable restrictions (free motion,
 harmonic attraction) \it  classical \rm  deterministic   phase-space  paths are
 supported by the associated with $\psi(x,t)$  positive    Wigner distribution function
 and its spatial  marginal distribution, c.f. \cite{gar} for a related discussion.

 However,  none of the above  \it   derived \rm  trajectory  "pictures" deserves the status of
 an underlying physical "reality" for quantum phenomena,  although  each of them  may serve
 as more or less  adequate  pictorial description of the  wave-packet dynamics,
 \cite{holland,rabitz}.

 It is in view of   Born's  statistical interpretation postulate, that the
  the Schr\"{o}dinger picture  dynamics  sets  a well  defined   transport problem for
   a probability density  $\rho (x,t) \doteq |\psi  (x,t)|^2$ which   one is   tempted to resolve
 in terms  of stochastic processes  and their  sample paths.
 A  direct  interpretation in terms of random   "trajectories"  of  a Markovian diffusion-type process is here
 in principle possible  under a number of mathematical restrictions, but may happen to be
non-unique and not necessarily  global in time. The nontrivial boundary data,  like the
presence of  wave  function  nodes, create
 additional problems  although the nodes  are known to be never reached by the pertinent processes.
 The main source of difficulty lies in guaranteing the existence of a process  per se i.e. of the
well defined (and unique, if possible) Markovian  transition probability  density function, which
 in its full generality still remains a profound mathematical problem, \cite{eberle}.
 A  related issue of the global existence of Bohmian trajectories has been addressed in
 \cite{durr}.

Both stochastic and causal (Bohmian)  trajectory interpretations, need the solvability of
 the Schr\"{o}\-din\-ger equation, hence  conditions a), b) and c) of Section 2  must
 be respected. Accordingly, with  $\psi (t)$    belonging  to the domain of $\hat{H}$, we
 infer from the  formula (\ref{finite})  that the finite energy condition automatically
 follows.  This state of affairs  hardly one can interpret as "incomplete",
 formally or physically, on the basis of Ref.~\cite{hall}.

 There is  no doubt
  that states  \it not \rm in the domain of $\hat{H}$  are not amenable to a straightforward
  trajectory interpretation, but this  feature was rather obvious from  the outset, in
  rigorous formulations of the pertinent theories, \cite{carlen,durr}.
   The real point is whether those  "outer"  states can be termed "physical", i. e. compatible
   with  well defined experimental procedures and their mathematical (quantum mechanical)  imaging,
   see  e.g.  Refs.~\cite{rauch,zeilinger}. This  issue has  been left untouched  in ref.~\cite{hall}.

{\bf  Acknowledgement:}
While preparing the present note I have  benefited from  discussions with
  Professor Witold Karwowski.  I convey my warm thanks to him for all  comments.
The paper has been supported by the Polish Ministry of Scientific
Research and  Information Technology under the (solicited) grant
No PBZ-MIN-008/P03/2003.

\end{document}